# Toward a Methodological Knowledge for Service-Oriented Development Based on OPEN Meta-Model

**Mahdi Fahmideh, Fereidoon Shams**

Automated Software Engineering Research Group, ECE Faculty, SB University GC, Tehran, Iran
{m_fahmideh, f_shams} @sbu.ac.ir

**Abstract.** Situational method engineering uses a repository of reusable method fragments that are derived from existing software development methodologies and industrial best practices to simplify the construction of any project-specific software development methodology aligned with specific characteristics of a project at hand. In this respect, OPEN is a well-established, standardized and popular approach for situational method engineering. It has a large repository of reusable method fragments called OPF that method engineers can select and assemble them according to the requirements of a project to construct a new project-specific software development methodology. In this position paper, we present the basic concepts and foundations of OPEN and argue for an urgent need for new extensions to OPEN and its repository in support of service-oriented software development practices.

## 1 Introduction

It has proven untenable that there is no universal Software Development Methodology (SDM) appropriate for all situations [1,2,3]. This consensus is due to the situation factors of projects at hand such as organizational maturity and culture, people skills, commercial and development strategies, business constraints, and tools issues. Software development organizations need to develop their own project-specific SDM for their software projects. In this respect, Method Engineering (ME) is an approach in which a project-specific SDM is constructed. The most well-known offered sub-set of ME for tailoring SDMs is called Situational Method Engineering (SME) [4,5,6] wherein a project-specific SDM is constructed from the assembly of a number of reusable Method Fragments [7] or Method Chunks [8] that are stored in a Repository or Method-Base [9,10]. Indeed the repository of method fragments is a Methodological Knowledge-Base that stores the knowledge about what and how to develop a software system [11]. Each method fragment bears a piece of knowledge for software development that is characterized by a name and an intention specifying the goal of the method fragment. More specifically, a method fragment can be thought of as a couple of two interrelated parts: product model and process model. The product part of a method defines a set of concepts and relationships between these concepts. In contrast, the process part describes how to construct the corresponding product part. Fig. 1 depicts a typical method fragment. This method fragment aims to provide a use-case model as a solution to resolve a problem description. The product part represents the required products and the process part provides suitable guidelines to make a use-case model.

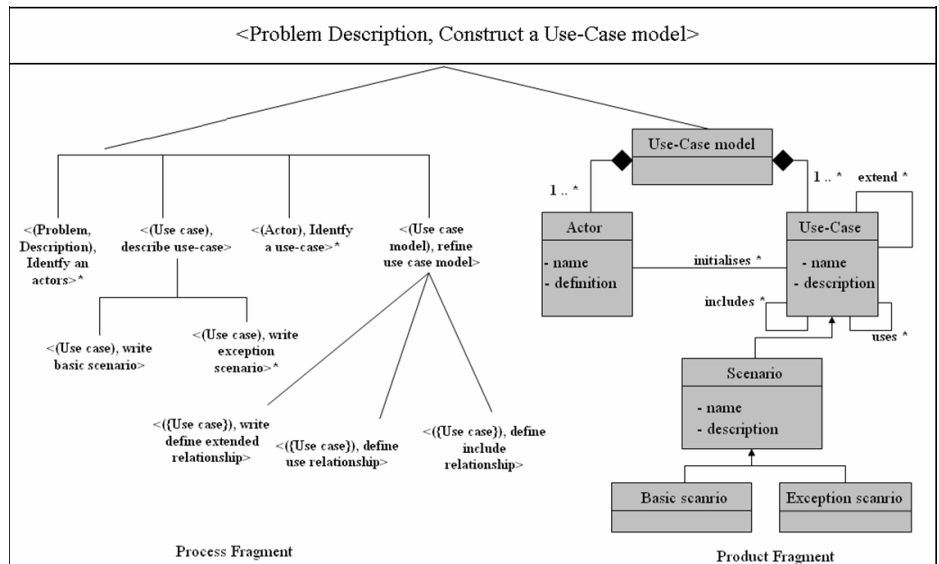

Fig.1. A method fragment (adopted from [11])

The Object-oriented Process, Environment, and Notation (OPEN) or Open Process Framework (OPF) [12] is a framework that is highly compatible with the ideas of SME approach. OPEN has a large number of method fragments stored in a repository, called OPF. Indeed the OPF repository is a methodological knowledge-base. In OPEN, methodological knowledge is represented in the OPEN meta-model format as well as method fragments. A method engineer can select and assemble method fragments and construct a project-specific SDM based on the unique set of characteristics of the project at hand. However, OPF contains method fragments mainly intended for Object-Oriented (OO) software development.

Given that newer approaches to software development have emerged that are currently practiced widely, there is an urgent need to enhance the OPF repository with new method fragments in support of these new approaches to software development [10]. One of these new and popular approaches is Service-Oriented Computing (SOC) that has been proposed in the last few years and is getting huge interest from software engineering researchers and practitioners. As stated in [13,14,15], SOC is a new computing approach that utilizes software services as the fundamental building blocks to facilitate the development of rapid, low-cost and easy composition of loosely-coupled fully distributed systems such as Clouds. A strand of inclination in the SOC field is the Service-Oriented (SO) software development subfield that grapples with development of SO software systems [15]. It should be noted it is needed to apply

SME idea approach to the SO software development context [16,17,18] because (1) the development of SO software are increasingly decentralized, (2) SO software is composed dynamically out of parts that are developed and operated by independent parties, (3) changes in requirements ask for continuous SO software adaptation and evolution, and that (4) the infrastructures on which SO applications run are fully distributed. This situation fosters the need for adoption of SME approaches in order to satisfy the various and changing requirements of SO software development.

Although OPEN has a large repository in support of construction of various types of software such as OO and Component-Based (CB) systems, it lacks any support for the development of SO systems. It is thus reasonable for the OPF repository to be extended to provide support for SO in addition to its current support for OO and CB software development. Therefore, in this paper we take a deep look at the foundations and basic constituents of OPEN and argue for extending OPF with method fragments in support of SO software development.

We have organized the rest of the paper as follows. Section 2 presents a brief overview of SME and OPEN. Section 3 argues in favor of adding SO extensions to the OPF repository. Section 4 concludes the paper with proposals for future work.

## 2 Overview

### 2.1 Situational Method Engineering

The prevalent belief that no single SDM can be applicable to all situations is the main reason for the emergence of ME [19]. Each software project has different characteristics so that a SDM tailoring should be adopted before starting the project. The ME approach was first introduced by Kumar [20] as a software engineering discipline aimed at constructing project-specific SDMs to meet organizational characteristics and projects situations. Brinkkemper [21] elaborated the definition of ME later as: "the engineering discipline to design, construct, and adapt methods, techniques and tools for the development of information systems". The most well known subset of ME is SME that is concerned with the construction, adaptation and enhancement of suitable SDM for the project at hand instead of looking for universal or widely applicable ones [21]. Based on the SME approach, a SDM is constructed from a number of encapsulated method fragments that have been already stored in a repository. The method fragments are the atomic elements of any SDM that a method engineer extracts from existing SDMs or from industrial best practices [22,23]. Method fragments are selected in such way to satisfy target SDM's requirements.

To realize SME, researchers have proposed many approaches [19]. Typically, the steps below are followed to construct a project-specific SDM:
1. Method engineer elicits and specifies the requirements of the target SDM based on the characteristics of the project at hand.
2. Then, he/she selects a number of most relevant method fragments from the repository based on a number of factors highly specific to the particular software development organization and particular situation of the project.
3. Method fragments are assembled to construct a full project-specific SDM.
4. To ensure high quality of the constructed SDM, a list of assessment criteria such as the ones proposed by Brinkkemper is used [24].

SME has been extensively used for OO software development [25]. One instance of the SME approach that is highly compatible and fits well with the above steps is extensively used for the development of a wide range of software projects, especially in the OO context is called OPEN [12]. OPEN defines a process meta-model that allows the elements of the OPEN, i.e method fragments, to be represented and reused. A large number of method fragments, stored in a single repository, called OPF repository, facilitates the instantiation of any project-specific SDM from the OPEN. The method fragments are reusable building blocks that can be adopted in more than one SDM construction effort [26]. Predefined rules and construction guidelines assist method engineers to select from repository and to assemble them. The instantiated SDM is mainly a new configuration of the OPEN. Successful industrial use of OPEN demonstrates its viability to software development [27]. This utilization in real world practices was the main reason we were motivated to extend OEPN with SO support.

**2.2 OPEN Process Framework**

OPEN or OPF is the oldest established SDM introduced in 1996 in an effort to integrate four SDMs namely, MOSES, SOMA, Synthesis and Firesmith [12,28]. OPEN is known as a popular SDM with full iterative-incremental lifecycle and process-focused SDM that is recently updated to become conformant with ISO/IEC 24744 [29]. It is mainly intended for use in either the development of a wide range of software systems or the construction of a wide-range of project-specific SDMs. OPEN is maintained by a not-for-profit Consortium consisting of an international group of methodologists, academics and CASE tool vendors [30]. As shown in Fig.2, OPEN contains an underpinning process meta-model, a single rich repository of method fragments, supportive tools, and usage guidelines that explain how method engineers can deploy the method fragments.

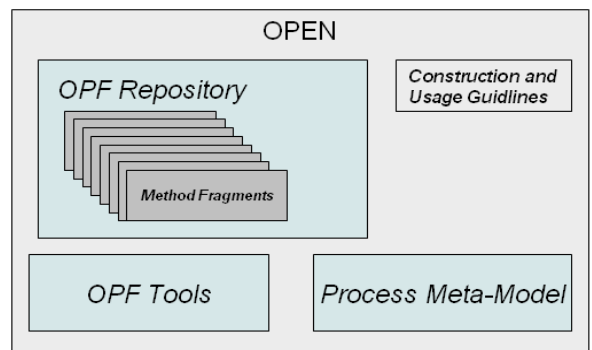

Fig.2. The main elements of the OPEN Process Framework (adopted from [12])

The OPEN's process meta-model provide a clear way to formally represent any method fragments e.g. process models, phases, activities, tasks, techniques, work products and roles. It is imperative that method fragments conform to the OPEN meta-model standard. This implies that new method fragments to be added to the repository must be conformant to this meta-model as well. The OPEN meta-model as shown in Fig.3 contains five core classes of method fragments as defined by Firesmith and Henderson-Sellers [12,30]:
1. **Work Unit**: Operations should be performed by Producer(s) or tools to develop required Work Products. Work Units based on their granularities are categorized in three levels of abstractions:
   - **Activity**: Some refer to Activity as software engineering discipline too. An activity is a coarse-grain type of typical Work Unit consisting of a cohesive collection of Tasks that produce a related set of Work Products. In other words, an Activity includes a group of relevant Tasks.
   - **Task:** A Task is a fine-grain type of Work Unit consisting of a cohesive collection of steps that produce Work Product(s).
   - **Technique:** An explicit procedure(s) that explains how a Task should be performed is called a Technique.
2. **Work Product**: Work Product is any significant produced artifact such as diagram, graphical and textual description, or program that is produced during software development.
3. **Producer**: Person(s) or tools that develop expected Work Products are a kind of Producer.
4. **Language**: Language represents the produced artifacts using a modeling language such as Unified Modeling Language (UML) [31] or any implementation language.
5. **Stage**: Stage is used for defining the overall macro-scale and time-box on a set of cohesive Work Units during the enactment of an instantiated OPEN. The instantiated process is structured temporally using the Stage concept element.

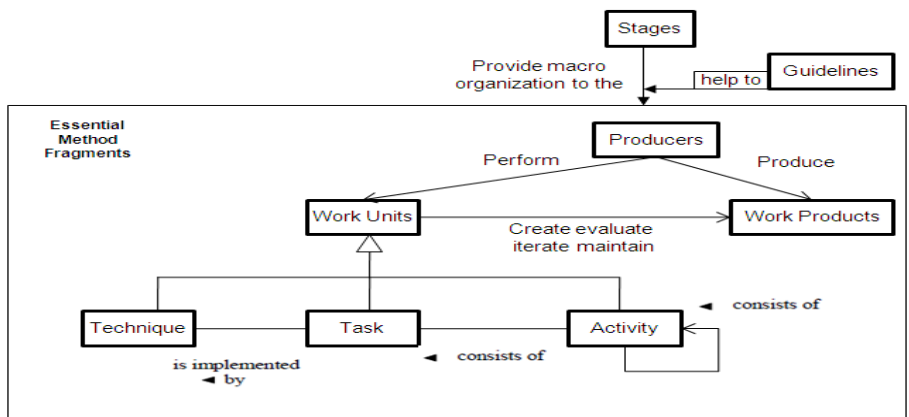

Fig.3. Constituents of the OPEN Process Meta-Model [12]

In addition to the process meta-model, OPEN contains a large number of method fragments at different levels of granularity (Activities, Tasks and Techniques) stored in a repository as it is shown in Fig.3. The process meta-model and repository of method fragments provide the underpinning and scaffolding context for situational method engineering. The OPF repository provides reusable method fragments as well as well-known and traditional activities for the construction of project-specific SDMs that are mainly intended for OO software development [12,30]. For instance, there are many tasks and techniques for Requirements Engineering such as requirements elicitation, use-case modeling, use-case specification, and prototyping.

There are many other approaches to development of software projects other than OO such as Component-Based Software Engineering (CBSE) or Component-Based Development (CBD). Although there is a lot of commonality between the traditional OO software development and other approaches of software development, the latter differs slightly from OO software development. On the other hand, we know that the most critical pre-requisite for construction of SDM is a repository that should be consistence with paradigmatic approaches [25,32]. Therefore, it is necessary that new method fragments are added to the OPF repository in support of new approaches of software development or technologies. Fortunately, the addition of new method fragments does not require any modifications to the underpinning OPF meta-model that has been standardized [25].

## 3 Discussion

Over the past years, several researchers have attempted to provide methodological knowledge to the OPF repository in support of various software development approaches. Henderson-Sellers et al have carried out the most sound and significant extension to the OPF repository. They have added many supportive method fragments to facilitate situational SDM construction in different approaches of software development such as:

- Extension Support for CBD [33].
- Extension Support for Web-Based Software Development [34,35].
- Extension Support for Aspect-Oriented Programming (AOP) [32].
- Extension Security-Related method fragments for the OPF [25].
- Several additional extensions in support of organizational transition and usage-centered design [36,37,38].

Therefore, OPF has matured with the introduction of method fragments for various approaches of software development. Recently, SOC has become more and more popular [13,14,15]. SOC has many favorable attributes including agility, reusability and standardization in a technical and in a business-oriented sense. In this respect, SO software development has received wide acceptance among practitioners and academia. Mainly, SO software development is considered as the next step towards resolving the deficiencies of CBD approach such as 1) the lake of standard interface that has been very difficult for software developers for component interoperability [15] and 2) ignorance of security issues of software components [39]. In the SO context, standard interfaces provide greater interoperability between service providers and consumers. Specifically SO is considered as an evolution of CDB software development [40]. In the context of SO, services are defined as reusable platform-independent building blocks of the system (or software). The standard interfaces provide greater interoperability between service providers and consumers and simplify the development of loosely coupled distributed systems [13]. This approach of software development is known as Service-Oriented Architecture (SOA), Service-

Oriented System Development, or Service-Oriented Software Engineering (SOSE). Since the SO approach can significantly improve the way software systems are developed, there has been an increase in the tendency for developing SO systems [15].

SO software development resembles the traditional waterfall process model and activities such as project planning, use-case modeling, OO Analysis and Design, Implementation and Test. However, many research publications and empirical evidences [15,41,42] report that the development of SO systems is different from the traditional software development. The SO software development has more challenges than traditional software development. Typically, an SO software development constituents activities such as service governance, service identification, specification and realization, service discovery and composition and service monitoring [14,43]. It is increasingly being recognized that modifications to traditional process models to suite the SO development, and introducing new software engineering activities and skills other than traditional activities for SO systems, are required. In addition, the development of SO systems need to apply SME approach to the SO software development in which a project specific SDM should be tailored specifically to meet the requirements of such projects. We strongly believe that there is a need to provide a repository we call it a Methodological Knowledge-Base that is extracted from successful experiences and best practices of SO development, for exchanging methodological knowledge among practitioners and software development organizations. In this context, similar endeavors have been accommodated in the area of requirements engineering [44], business interoperability [45] and method engineering [46]. For instance, the [44] proposed a set of classified and formalized patterns stored in a repository. The patterns represent recurrent problems and solutions during requirements engineering that could be adopted by developers.

An essential need to have such methodological knowledge so that it would be as much as useful is its representation. The SO methodological knowledge should be well-documented and maintained in a well-structured format so that one can easily understand and utilize it in real projects. Integrating different kinds of SO methodological knowledge in a common repository, such as OPF method fragments, has the following benefits:

- **Documentation**: Provides software development organizations with well-documented and well-structured useful knowledge of SO development.
- **Reusability**: The repository can be very useful to construct SO SDMs by assembling existing fragments or to adapt existing SDMs by adding specific fragments.
- **Continuous Evolution**: Getting feedbacks from practitioners, analyzing the feedbacks, and keeping methodological knowledge alive and up to date. Moreover, it is open and allows new practitioners to contribute new method fragments into repository.
- **Share knowledge**: To achieve this aim, OPEN is a good candidate because OPEN provides a standard meta-model for representation of methodological knowledge via autonomous and coherent method fragments. Moreover, OPEN provides good support from various approaches of software development and ideas of SME. The methodological knowledge provides support to software development organizations to help them construct project-specific SDM and share knowledge of developing SO systems with other practitioners.

In spite of full support of OPEN for various software development approaches, we have identified a deficiency in the current OPF. To the best of our knowledge, there is no support and similar reported research into defining specific method fragments for SO development. We advertise in support of methodological knowledge for SO development by providing a set of method fragments. These fragments are stored in the OPF repository alongside pre-existing method fragments. The applicability of the extended OPF is that any method engineer can construct a new SDM by selecting from existing method fragments for traditional activities and selecting from SO specific method fragments for SO development.

To achieve a methodological knowledge for development of SO systems in the format of OPEN method fragments, the source in which knowledge is extracted from is the main prerequisite. As stated in [47], one way to construct new method fragments is to utilize the existing SDMs as a source called existing method re-engineering. In this approach, a SDM is fully decomposed into a number of method fragments ready to be stored in the repository. To realize this approach, we have developed a manual procedure for identifying reusable method fragments from SDMs [48]. The procedure gets a SDM and proposes several steps to the method engineer to decompose SDM into a set of method fragments. The constructed method fragments are represented in the OPEN meta-model standard (as mentioned in Section 2.2) which is deemed to enhance the OPF repository so that they can be immediately imported into OPF tools supports. Indeed, the constructed method fragments will be methodological knowledge in SO development approach. As shown in Fig 4, the new method fragments are placed along with other existing method fragments as well.

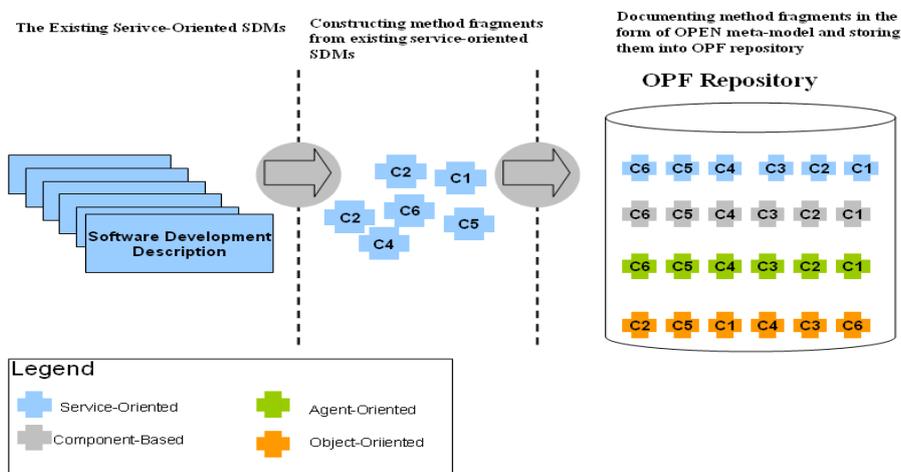

Fig.4. The overall process of constructing new SO specific method fragments

## 4  Future Work

We have started a thorough study of the fundamental issues in SO development, specifically in the current prominent SO SDMs, and have short-listed ten candidates. The candidates are IBM SOMA 2008 [43], SUN SOA Repeatable Quality (RQ) [49], CBDI-SAE Process [50], MSOAM [51], IBM RUP for SOA [52], Methodology by Papazoglou [14], IBM SOAD [53], SOUP [54], Steve Jones' Service Architectures [55] and Service-Oriented Architecture Framework [56]. These SDMs prescribe successive systematic activities in order to fulfill SO issues. These SDMs have been selected because of their empirical evidence, higher rate of citations, more accessible resources, and better documentations. In future, we intend to derive the commonalities between these SDMs and propose a set of method fragments. The method fragments convey from OPEN meta-model so they can easily be added to OPF repository. We envisage three levels of granularities for these specific SO method fragments:

- Activity: Some of the existing method fragments for Activities in the OPF would be enhanced by the incorporation of SO ideas specific to SO Task method fragments.
- Tasks: New SO Task method fragments.
- Techniques: For each Task method fragment, a number of supportive Techniques will be provided.

Therefore, future research offers the new SO specific method fragments for OPF repository.